\documentclass[structabstract]{aa}
\usepackage{graphicx}
\usepackage{txfonts}
\usepackage{amssymb}
\usepackage{multirow}
\makeatletter
    
    \newcommand{\Rmnum}[1]{\expandafter\@slowromancap\romannumeral #1@}
\makeatother

\begin{document}

%
%
   \title{Constructing a refined model of small bodies in the solar system -- \Rmnum1. The Jovian Trojans}

   \author{J. Li
           \and
           Y.-S. Sun
          }

   \institute{School of Astronomy and Space Science \& Key Laboratory of Modern Astronomy 
                     and Astrophysics in Ministry of Education, Nanjing University, 163 Xianlin Road, Nanjing 210023, PR China\\
    \email{ljian@nju.edu.cn}
             }

   \date{\today}

  \abstract
   {}
   {We construct a new arc model to represent the global perturbation induced by the Jupiter Trojans for the development of the modern planetary ephemerides.}
   {The Jupiter Trojans are divided into two groups: (1) the 226 biggest ones with absolute magnitudes $H<11$, have well determined masses and are treated as individual bodies; (2) the remaining small ones with $H\geqslant11$, are modeled by two discrete arcs centered at Jupiter's L4 and L5 points, respectively. Then we derived the parameters characterizing the arcs, such as the total mass, the mass ratio, and the spatial distributions. Uncertainties on the fitted parameters have also been taken into account.}
   {We find that the total mass of all the Jupiter Trojans, including the undiscovered ones, is most likely about $1.861\times 10^{-5}$ Earth mass. Then the global perturbation of the trojan population induced on the Earth-Mars distance is numerically estimated to be about 70 m during the 2014--2114 time interval. We also give a simple analytic expression. For the satellites of Mars and Jupiter, however, the change of the distance between the Earth and a satellite can be much more significant, reaching on tens of thousands of meters over one century, due to perturbations of the Jupiter Trojans.}
   {}


   \maketitle
%

\section{Introduction}

At present, computational celestial mechanics has achieved very high precision through computer-assisted techniques. Nevertheless, when predicting positions of celestial bodies in the solar system by using planetary ephemerides, there is still somewhat of a discrepancy with direct space observations. For Mars, the uncertainty of today's ranging measurements from spacecraft (e.g., Mars Odyssey and MRO) is typically only 1 m (Konopliv et al. 2011), while modern ephemerides are seen to extrapolate its position one year into the future with an accuracy of about 15 m (Folkner et al. 2008; Fienga et al. 2010).  The ephemeris error, either for the short-term space mission or the long-term chaotic evolution of the solar system, could be crucial. To construct an appropriate dynamical model of the motion of celestial bodies, it is necessary to take into account the perturbation of numerous asteroids (e.g., main belt asteroids, Jupiter Trojans, Kuiper belt objects, and others),  which is a contributing factor that can not be ignored.

The main belt perturbation has been deeply explored in many previous works (Krasinsky et al. 2002; Konopliv et al. 2006; Kuchynka et al. 2010; Fienga et al. 2011; Pitjeva \& Pitjev 2014). The general approach to model the total influence of the main belt asteroids uses a homogeneous ring to represent the numerous small asteroids, other than the 300 or so largest ones
(e.g., Ceres, Pallas, Juno), which are considered individually. By comparing the numerical ephemeris of Mars with the high-accuracy ranging data from martian spacecrafts, the two parameters (mass and radius) characterizing the ring could be derived. Consequently, the planet motion could be improved based on such a ``ring'' model and a better fit to observations achieved. According to the evaluation from the ephemeris INPOP06 (Fienga et al. 2008), the effect of the asteroidal ring on the perturbation of the Earth-Mars distance reaches approximately 150 m during the 1969--2010 time interval. However, The perturbation from the main belt asteroids is not sufficiently accurate, leading to  somewhat divergence in the ephemerides of Mars as time passes. It has been shown that this is due to the uncertainty of knowing the mass values of many of the largest asteroids (Pitjeva \& Pitjev 2014).

Yet, there is another group of small bodies that essentially share the orbit of Jupiter, but lead or trail this planet by about 60$^\circ$ of longitude, i.e., around the L4 or L5 triangular Lagrangian point. These objects are known as Jupiter Trojans, and said to be trapped in Jupiter's 1:1 mean motion resonance. At the time of writing, more than 6000 Jupiter Trojans have been registered in the Minor Planet Center\footnote{http://www.minorplanetcenter.net/iau/lists/JupiterTrojans.html} (MPC), and they also could affect the motion of planets, especially for Mars. However, differing from the main belt asteroids, the Jupiter Trojans are distributed in two discrete regions, which are co-orbiting the Sun ahead of and behind Jupiter, respectively. Therefore, the gravitational perturbations induced by these two swarms can not be simply mimicked by a homogeneous ring. The other two obstacles to properly model the Jupiter Trojans are the uncertainties of their total mass and the number asymmetry between the L4 and L5 swarms. According to Jewitt et al. (2000) and Fern\'andez et al. (2009), there are $\sim1.6 \times 10^5$ Jupiter Trojans with diameters $D > 2$ km populated around the L4 point, and their combined mass is on the order of $10^{-4} M_{\oplus}$ ($M_{\oplus}$ is the Earth mass); while the number of trojans occupying the L5 region could be relatively smaller. The number ratio for these two swarms, denoted by $N(\mbox{L}4)/N(\mbox{L}5)$ hereafter, currently is not deterministic and may vary between 1.3 and 2 (Szab\'o et al. 2007; Grav et al. 2011; Grav et al. 2012; Nesvorny et al. 2013; Di Sisto et al. 2014).  In addition,we hardly have any information about the Jupiter Trojans with sizes smaller than $D \sim 1-2$ km, while this population could reach a total number up to millions.  

As we know, in development of planet ephemerides so far, the perturbations of the Jupiter Trojans have not been taken into account. We believe that estimations need to be made for the effects of Jupiter Trojans on the motion of the planets or satellites, and this work is clearly warranted for refining the current dynamical model of the solar system. In the future, if reconciling the results obtained in this paper with new data from the on-going space missions like InSight and Lucy, it is possible to help us to further constrain the physical parameters of Jupiter Trojans, in particular, their total mass which has the largest uncertainty.

The rest of this paper has been divided into the following sections. In Sect. 2 we estimate the masses of all the Jupiter Trojans, including the hidden ones. Section 3 is devoted to constructing a new ``arc model'' to represent the global perturbation from the Jupiter Trojans, instead of integrating hundreds of thousands of individuals. Validation tests have been conducted for our arc model, and then the effect of the Jupiter Trojans on the Earth-Mars distance is carried out with reasonable physical parameters. In Sect. 4 we apply the arc model to evaluate the perturbations from the Jupiter Trojans on the motion of natural or artificial satellites around Mars and Jupiter. Our conclusions are presented in Sect. 5.


\section{Estimations of Jupiter Trojan masses}

As the first step to model the global perturbation from a large number of Jupiter Trojans, we describe in this section an analytical approach to the mass determination, even including those unseen small trojans. We have adopted 6064 Jupiter Trojans from the MPC as our nominal samples. About 30\% of them have accurately measured diameters through \textit{WISE/NEOWISE} observations (Grav et al. 2011; Grav et al. 2012). 

In the process of estimating the masses of Jupiter Trojans, we split them into two groups at a specific absolute magnitude of $H=11$: \\

(1) For the 226 known Jupiter Trojans with $H<11$, this sample is believed  to be complete since no objects brighter than the limit of $H_{\mbox{\scriptsize{complete}}}=12.3$ have been discovered since mid-2006; and, they have available \textit{WISE/NEOWISE} diameters $D$ and geometric albedos $p_v$. Here and henceforth, we set a default bulk density of $\rho=2000$ kg/m$^{-3}$ for any Jupiter Trojans (Jewitt et al. 2000). Then we can simply derive a mass for each of these large objects. The total mass of this trojan group is thus estimated to be $\sim1.127\times10^{-5} M_{\oplus}$.\\

(2) For the smaller trojans with $H\geqslant11$, they are far from observational completeness, and most of them have no size measurements. Considering these two important problems, in the following we will present an updated/new differential size distribution of faint Jupiter Trojans fitted by current survey data, and then evaluate their reasonable masses.

Theoretically speaking, the small-body population with diameters below a certain  ``break'' value could reach a state of collisional equilibrium, and the differential size distribution would follow the power law
\begin{equation}
N(D) \propto D^{-q}, 
\label{diff_D_dist}
\end{equation}
where the slope should have evolved to a canonical value of $q=3$ for erosive collisions (Kenyon \& Bromley 2004, 2012). Since the translation from the diameter $D$ to the absolute magnitude $H$ can be made using the formula
\begin{equation}
H=19.14-2.5\log(p_v/0.04)-5\log(D), 
\label{H&D}
\end{equation}
if the albedo $p_v$ is size-independent, then the differential absolute magnitude distribution is described by
\begin{equation}
N(H)dH=N_0\times10^{\alpha H}dH, ~~~\alpha=(q-1)/5=0.4,
\label{diff_H_dist}
\end{equation}
where $N_0$ is a constant to be determined from observations.

\begin{figure}
  \centering
  \includegraphics[width=9.5cm]{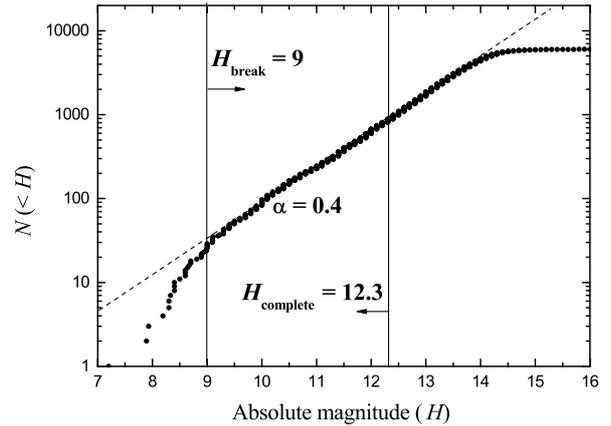}
  \caption{Cumulative absolute magnitude ($H$) distribution for the observed Jupiter Trojans. Trojans with $H\le H_{\mbox{\scriptsize{complete}}}=12.3$ are observationally complete. Trojans with $H\ge H_{\mbox{\scriptsize{break}}}=9$, are thought to have reached a state of collisional equilibrium, and the differential $H$ distribution can be well described by $N(H) \propto 10^{\alpha H}$, with the slope $\alpha=0.4$. The number dropping beyond $H=14$ is probably due to undiscovered Jupiter Trojans at fainter magnitudes.}
  \label{fig:Hdistribution}
\end{figure}

Figure \ref{fig:Hdistribution} shows the cumulative absolute magnitude distribution for our nominal samples from the MPC catalog. We note that, beyond the break at $H_{\mbox{\scriptsize{break}}}=9$ (equivalent to $D_{\mbox{\scriptsize{break}}}\approx21\sqrt{p_v}$), the slope of the best-fit line (dashed) for the power law is $\alpha=0.4$. This result is consistent with both the above argument and the analysis done in Morbidelli et al. (2009). However, for the population that has $H$ exceeding a value of about 14 ($>H_{\mbox{\scriptsize{complete}}}$), the slope becomes shallower. The reason for the number drop-off is apparently related to the incompleteness of today's observation. Thus, we can assume that the slope $\alpha$ of the absolute magnitude distribution (\ref{diff_H_dist}) is held constant to the faint end of Jupiter Trojans with $H\gg H_{\mbox{\scriptsize{complete}}}$.

Next, we would like to derive the value of the constant $N_0$. From Eq. (\ref{diff_H_dist}), the cumulative number of Jupiter Trojans with $H_1\leq H \leq H_2$ can be written as
\begin{equation}
N(H_1 \leqslant H \leqslant H_2)=\!\!\!\int_{H_1}^{H_2}\!\!N_0\times10^{\alpha H}dH=N_0\frac{10^{\alpha H_2}-10^{\alpha H_1}}{\alpha\ln10},
\label{H_dist}
\end{equation}
where $H_1 > H_{\mbox{\scriptsize{break}}}$. Since the subset of trojans with $9.1 \leq H \leq 12.3$ is complete, and the corresponding observed number is 778, we can determine $N_0\approx  1.0023 \times 10^{-2}$. This outcome implies that the total population of Jupiter Trojans with $12.3 < H \leq 15$ is approximately ten thousand, while currently only about a half of them have been reported to the MPC. 

After fitting the absolute magnitude distribution in Eq. (\ref{diff_H_dist}) available for $H\rightarrow\infty$, we now have the expression for the mass distribution of the considered trojan group with $H\geqslant11$, which reads
\begin{equation}
M(H\geqslant11)=\int_{H=11}^{\infty}\frac{4}{3}\pi\rho\left( \frac{D}{2}\right)^3N(H)dH,
\label{Mtotal_1}
\end{equation}
where the bulk density $\rho$ is the same as that of large trojans with $H<11$. With Eq. (\ref{H&D}) for the relationship between $D$ and $H$, we can get the analytical expression of the integral (\ref{Mtotal_1})
\begin{equation}
\frac{M(H\geqslant11)}{1\mbox{kg}}=\frac{5.56\times10^{19}}{(p_v)^{3/2}}\cdot10^{-0.2H},~~~H=11.
\label{Mtotal_2}
\end{equation}
Given a typical albedo of $p_v=0.04$ (Jewitt et al. 2000; Fern\'andez et al. 2003; Nakamura \& Yoshida 2008; Yoshida \& Nakamura 2008), which is regarded as the mean value for Jupiter Trojans and/or primitive small bodies in the outer solar system, we obtain from Eq. (\ref{Mtotal_2}) that the total mass of Jupiter Trojans with $H\geqslant11$ is $\sim7.343\times10^{-6}M_{\oplus}$.\\

Finally, including the masses of 226 largest trojans with $H<11$ as acquired before, we can estimate the total mass of the Jupiter Trojans to be
\begin{equation}
M_{\mbox{\scriptsize{tot}}}=M(H<11)+M(H\geqslant11) \approx 1.861\times 10^{-5}M_{\oplus},
\label{M_JTs}
\end{equation}
which includes the hidden mass. The value $M_{\mbox{\scriptsize{tot}}}$ is about 4.6\% of the total mass of the main belt asteroids, which is $\sim4.079\times10^{-4}M_{\oplus}$ provided by Pitjeva \& Pitjev (2014). For comparison, we also calculate the total mass of 6064 nominal samples from the MPC, denoted by $M_{\mbox{\scriptsize{tot}}}^{\mbox{\scriptsize{MPC}}}$. Among these real trojans, about 30\% have diameters available from \textit{WISE/NEOWISE} data; for the remaining objects, their diameters are computed from the assumed $p_v=0.04$ and the observed $H$ by Eq. (\ref{H&D}). This approach yields a value of  $M_{\mbox{\scriptsize{tot}}}^{\mbox{\scriptsize{MPC}}}=1.497\times10^{-5}M_{\oplus}$, which agrees with the order of magnitude of the predicted $M_{\mbox{\scriptsize{tot}}}$ but is 20\% less. Obviously, the difference in the total mass is due to the observational incompleteness, for example, 99\% of Jupiter Trojans smaller than the 1km scale remain unobserved (Jewitt et al. 2004). Our mass estimate is comparable to that in the work of Vinogradova \& Chernetenko (2015), which provided a total mass of the Jupiter Trojan at $(0.999\pm0.633)\times 10^{-5}M_{\oplus}$, but a bit larger. This difference may be partly due to the lower fraction of hidden components ($\sim7\%$) in their model.


In the framework of the Nice model, Morbidelli et al. (2005) numerically studied the capture of planetesimals into co-orbital motion with Jupiter in the early solar system, they predicted a mass of $0.4-3\times 10^{-5}M_{\oplus}$ for the trapped trojan population, of which the median value seems quite consistent with our analytic estimation. However, Morbidelli et al. (2005) assumed a bulk density of $\rho'=1300$ kg/m$^{-3}$, and a mean albedo of $p'_v=0.056$ in their calculations. If we revise our analysis corresponding to these two different parameters, the total mass of the Jupiter Trojans would reduce to $1.020 \times 10^{-5}M_{\oplus}$, which still falls in the range of the mass produced by Morbidelli et al.'s model.  

Besides the uncertainties on $\rho$ and $p_v$,  the detailed shape information is also lacking for the Jupiter Trojans. Here we simply assign each object a volume computed assuming a spherical shape, while a large proportion of observed samples actually have been found to be nonspherical (Grav et al. 2011). This means that the trojan mass $M_{\mbox{\scriptsize{tot}}}$ that we derived could be either higher or lower by some amount.

Despite these uncertainties, the mass obstacle has little impact on building our arc model for the Jupiter Trojans. In the following section, we would consider those masses used in Eq. (\ref{M_JTs}), either for each single trojan with $H<11$ or for the smaller size population with $H\geqslant11$, as standard values. And furthermore, the influence of different trojan mass would be investigated and then discussed.


\section{The arc model}

The unperturbed solar system consists of the Sun and eight planets from Mercury to Neptune. Then, the arc model is implemented to represent the exterior force induced by the Jupiter Trojans. The 226 largest trojans (hereafter referred to as Bigs) with well determined masses are treated as individuals, and they are included into the process of simultaneous numerical integration, while their mutual perturbations are neglected. The orbital elements of the Bigs from the MPC, and the positions and velocities of the planets from DE405 (Standish 1998), are adjusted to the heliocentric frame referred to the J2000.0 ecliptic plane at epoch 2014 May 23. 

For the remaining small trojans with $H\geqslant11$, they are represented by two discrete massive arcs centering at Jupiter's L4 and L5 points, respectively. The gravitational potential at $(x', y', z')$ caused by either of these two arcs can be described as
\begin{eqnarray}
  &&U(x',y',z',t)=-\int_{r}\int_{z}\int_{\lambda_J(t)+\phi_1}^{\lambda_J(t)+\phi_2}
                               \frac{G\rho_A r}{dist} dr dz d\phi,\nonumber\\
  &&dist=\sqrt{(r\cos\phi-x')^2+(r\sin\phi-y')^2+(z-z')^2},
\label{arc}
\end{eqnarray}
where $G$ is the gravitational constant, $\rho_A$ is the volume density of the arc, $(r, z, \phi)$ are the cylindrical coordinates with respect to the Sun, and $\lambda_J(t)$ is the mean longitude of Jupiter at the moment $t$, the phase angle between the leading (trailing) arc and Jupiter ranges from $\phi_1>0$ $(<0)$ to $\phi_2>0$ $(<0)$. We note that the potential $U$ is a function of $\lambda_J(t)$, and explicitly is time dependent. 

Numerically integrating Eq. (\ref{arc}) in the three-dimensional parameter space would be extremely computationally intensive, particularly as we aim to incorporate this procedure into the planetary ephemeris computing. Since the potential $U$ should be radially symmetric with respect to the mean orbit of Jupiter and also roughly symmetric in the horizontal direction, let us first consider the one-dimensional arcs having a uniform mass distribution in the azimuthal direction. For this simple case, in Eq. (\ref{arc}) we chose a set of standard parameters: $r=5.2$ AU, $z=0$, $\phi_1=30^{\circ}$ $(-90^{\circ}$) and $\phi_2=90^{\circ}$ $(-30^{\circ}$) for the L4 (L5) arc; the (linear) density $\rho_A$ is calculated from the standard mass of $M_{\mbox{\scriptsize{arcs}}}=M(H\geqslant11)$ determined in Sect. 2. The leading-to-trailing number ratio is chosen to be  $N[\mbox{L4}]/N[\mbox{L5}]=1.6$ (Szab\'o et al. 2007), which is equivalent to the mass ratio of the two arcs as the same bulk density $\rho$ is assumed for all the trojans. We refer to these two specific one-dimensional arcs as standard arcs for comparison purpose below.

\subsection{Gravitational perturbations on the Earth-Mars distance}

\begin{figure}
  \centering
  \includegraphics[width=9.5cm]{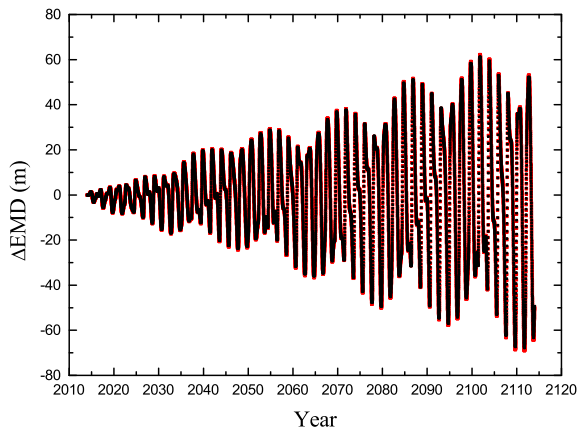}
  \caption{Perturbation on the Earth-Mars distance induced by the Jupiter Trojans in the time interval between years 2014 and 2114: (black) the combined effect of the 226 biggest trojans with $H<11$ (referred to as Bigs) together with the remaining small ones (modeled by two arcs), from a simultaneous integration; (red) the sum of individual contributions from the Bigs and the arcs, from two separate integrations.}
  \label{fig:EMD}
\end{figure}

In order to evaluate the gravitational effect of the Jupiter Trojans, as is usually done, we estimated the change in the Earth-Mars distance
\begin{equation}
\Delta EMD=EMD_1-EMD_0,
\label{Delta_EMD}
\end{equation}
where $EMD_1$ and $EMD_0$ indicate the Earth-Mars distances calculated by including or not trojan perturbations, respectively. The perturbed and unperturbed solar systems are then numerically integrated, using a 19th-order Cowell prediction-correction algorithm, on the 2014--2114 time interval. The calculations are controlled by the relative error in the trojan's Jacobi integral, which is set to be $<10^{-3}$. Constrained by this limit, the integration step size is adjusted to be 0.5 day. Moreover, in the arc model, a careful examination of all the 226 Bigs' trajectories over 100 years confirms that these objects are always locked in the 1:1 mean motion resonance with Jupiter. Figure \ref{fig:EMD} shows the perturbation on the Earth-Mars distance induced by the Bigs together with the two standard arcs in a simultaneous integration (in black). We find that the effect of the Jupiter Trojans could reach a value of $\Delta EMD\approx70$ m over one century. 

Alternatively, we estimated the individual perturbations induced by the Bigs and the arcs, denoted by $\Delta EMD_{\mbox{\scriptsize{Bigs}}}$ and  $\Delta EMD_{\mbox{\scriptsize{arcs}}}$, respectively. We were then able to approximate to the global effect caused by the Jupiter Trojans by the sum of $\Delta EMD_{\mbox{\scriptsize{Bigs}}}$ and  $\Delta EMD_{\mbox{\scriptsize{arcs}}}$, which is plotted with the red curve in Fig. \ref{fig:EMD}. Comparing this to the result from the combined case (black curve), we see that they coincide very well. 

To clarify this quantitative coherence, we can expand the perturbed Earth-Mars distance $EMD_1$ in a Taylor series in terms of the trojan mass $M_i$ (Kuchynka et al. 2010)
\begin{eqnarray}
  EMD_1(M_1, \cdots, M_{n})&\!\!\approx\!\!&EMD_0+\frac{\partial EMD_1}{\partial M_1}M_1+\cdots + \frac{\partial EMD_1}{\partial M_n}M_n,
\label{taylor1}
\end{eqnarray}
which is to first-order accuracy. As in the arc model only two contributions of perturbations are considered, one from the Bigs, and the other from the remaining small trojans represented by the arcs, the above expression can be rewritten as
\begin{equation}
  EMD_1\approx EMD_0+\sum_{i=1}^{226}\frac{\partial EMD_1}{\partial M_i}M_i~{\mbox{(for Bigs)}}+\frac{\partial EMD_1}{\partial M_{\mbox{\scriptsize{arcs}}}}M_{\mbox{\scriptsize{arcs}}}.  
\label{taylor2}
\end{equation}
Substituting Eq. (\ref{taylor2}) into Eq. (\ref{Delta_EMD}), we obtain
\begin{equation}
\Delta EMD\approx \Delta EMD_{\mbox{\scriptsize{Bigs}}} + \Delta EMD_{\mbox{\scriptsize{arcs}}},
\label{taylor3}
\end{equation}
where
\begin{eqnarray}
\Delta EMD_{\mbox{\scriptsize{Bigs}}}&\approx&\sum_{i=1}^{226}\frac{\partial EMD_1}{\partial M_i}M_i,\nonumber\\ 
\Delta EMD_{\mbox{\scriptsize{arcs}}}&\approx&\frac{\partial EMD_1}{\partial M_{\mbox{\scriptsize{arcs}}}}M_{\mbox{\scriptsize{arcs}}}.
\label{taylor4}
\end{eqnarray}
Equation (\ref{taylor3}) can roughly explain the high-degree overlap between red and black curves in Fig. \ref{fig:EMD}, from which we argue that it is reasonable to treat the Bigs and the arcs independently in the calculation of perturbations of the Jupiter Trojans.

For the sake of simplicity, we would like to quarantine the perturbations induced by the Bigs. The number of this population with $H<11$ is certain since they are observationally complete. If we have added the observational errors in the diameters, the upper limit sizes of the Bigs can be determined. Then we are able to assign each of them a maximum mass, and that would result in a total mass of only about 4\% heavier than the standard one. Using this slightly different set of masses for the Bigs, we find that the perturbation on the Earth-Mars distance is nearly the same as we obtained before, in other words, the maximum of $\Delta EMD_{\mbox{\scriptsize{Bigs}}}$ reaches around 30 m in one century. Therefore, $\Delta EMD_{\mbox{\scriptsize{Bigs}}}$ is to be considered as a fixed value, at a specific epoch. 

However, the residual perturbations induced by the smaller population with $H\geqslant11$ are fraught with large uncertainties due to the deficit of (individual and total) mass measurements. In the following we focus on developing a rather general approach based on the standard arcs that could allow us to better improve the arc model.

\subsection{Validity testing}

\begin{figure}
  \centering
  \begin{minipage}[c]{0.5\textwidth}
  \centering
  \hspace{0cm}
  \includegraphics[width=9.5cm]{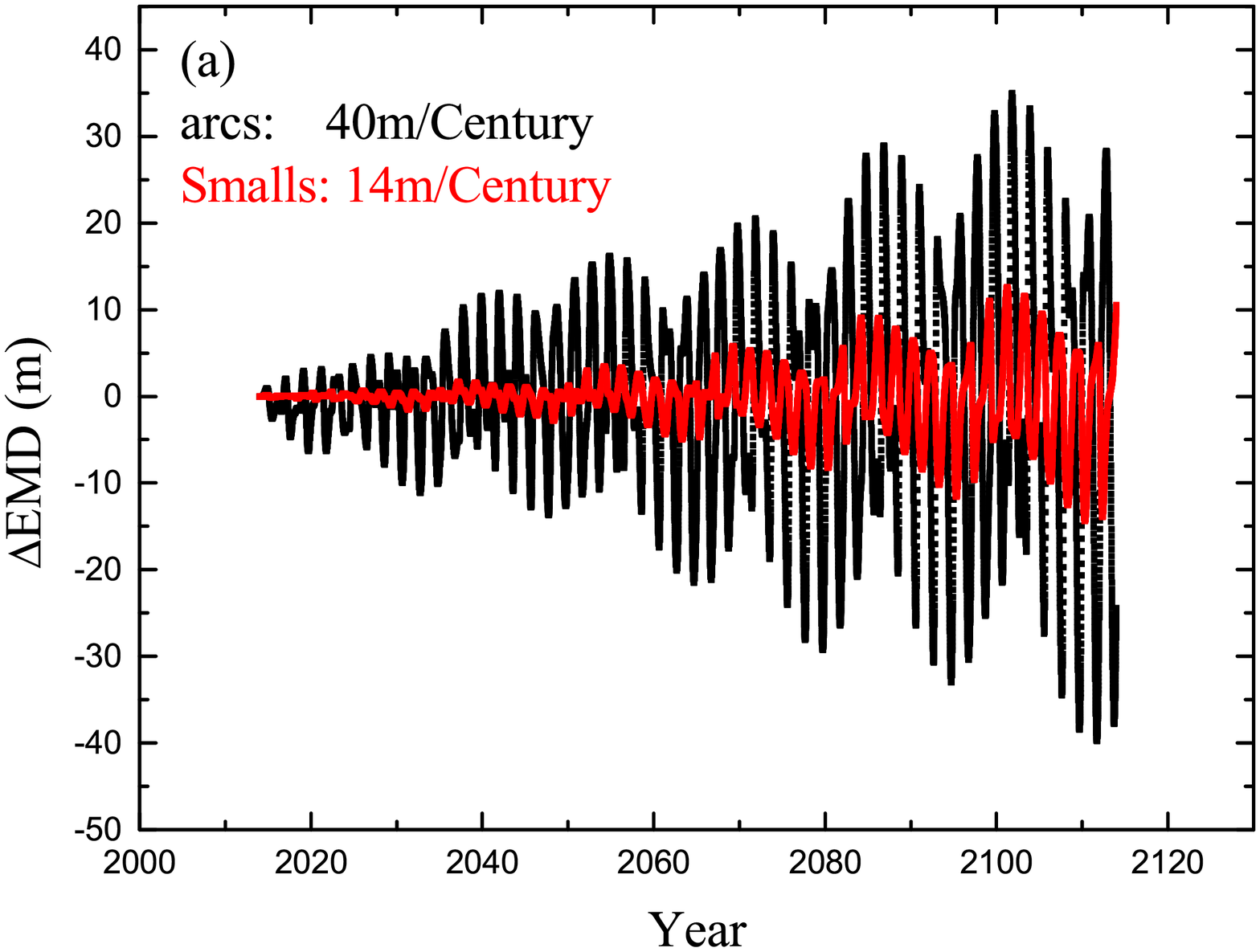}
  \end{minipage}
  \begin{minipage}[c]{0.5\textwidth}
  \centering
  \hspace{0cm}
  \includegraphics[width=9.5cm]{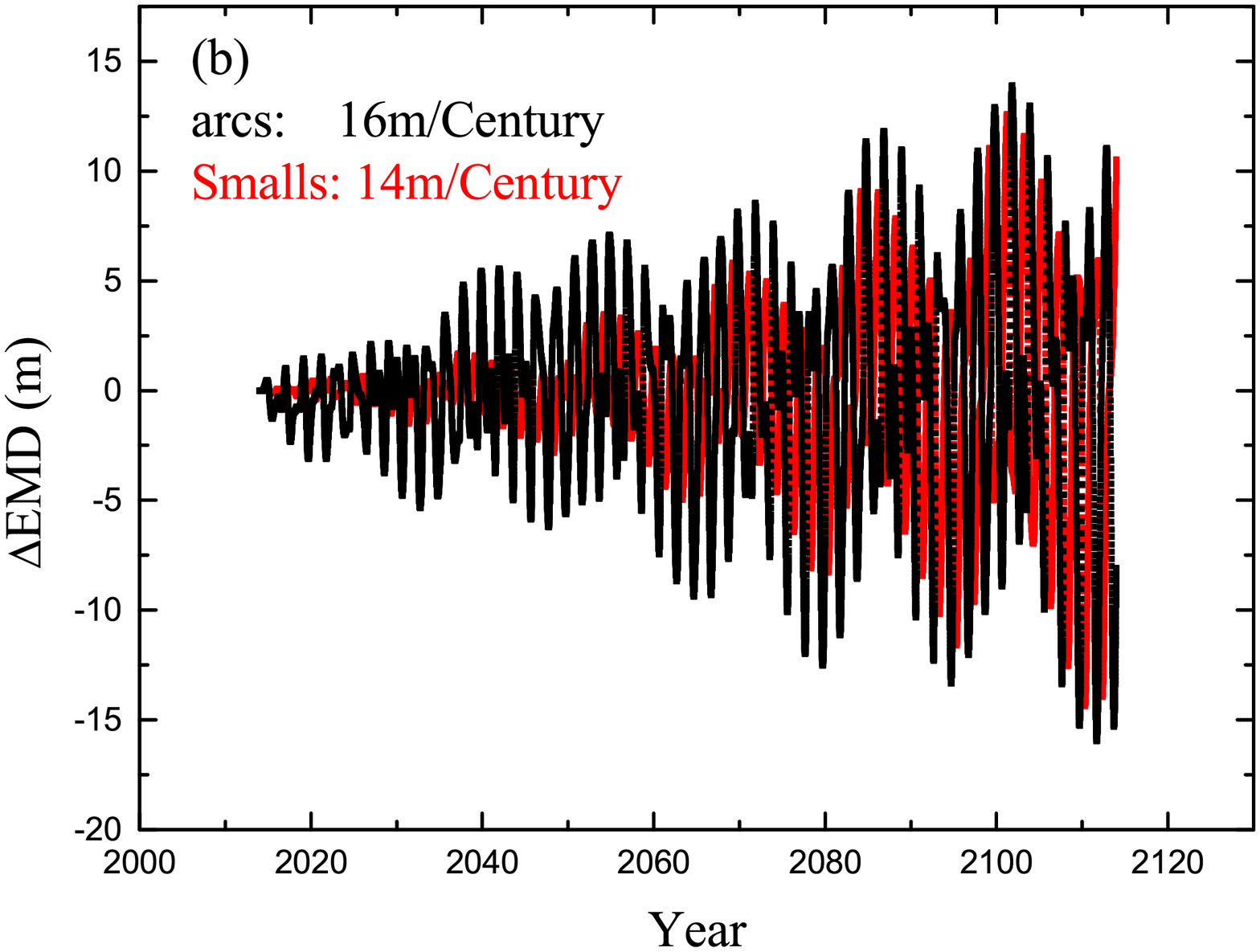}
  \end{minipage}
  \caption{Perturbation on the Earth-Mars distance induced by the two arcs (black) and the observed Jupiter Trojans with $H\ge11$ (referred to as Smalls) (red): (a) $M_{\mbox{\scriptsize{arcs}}}=7.343\times10^{-6}M_{\oplus}$, including the hidden mass of small trojans; (b) $M_{\mbox{\scriptsize{arcs}}}=M_{\mbox{\scriptsize{Smalls}}}=3.681\times10^{-6}M_{\oplus}$.}
 \label{fig:validation}
\end{figure}  

Naturally, one should first verify the ability of two discrete arcs to mimic a large number of Jupiter Trojans. We selected the 5838 observed trojans with $H\geqslant11$ (Smalls, hereafter) from our nominal samples. For these objects, we assigned their masses as in Sect. 2, and the global perturbation on the Earth-Mars distance is denoted by $\Delta EMD_{\mbox{\scriptsize{Smalls}}}$. We then performed direct N-body integrations, by adding or not adding the Smalls to the unperturbed solar system, to calculate the value of $\Delta EMD_{\mbox{\scriptsize{Smalls}}}$. 

Figure \ref{fig:validation}(a) shows the time evolution of $\Delta EMD_{\mbox{\scriptsize{Smalls}}}$ for the Smalls (red curve), as well as $\Delta EMD_{\mbox{\scriptsize{arcs}}}$ induced by the standard arcs obtained before (black curve). We find that these two curves do not coincide with each other, as the black peak is about three times the red one, meaning that, the perturbation from the arcs is stronger than that from individual Smalls. This can be understood by acknowledging the incompleteness of the faint Jupiter Trojans. The total mass of the Smalls is estimated to be $\sim3.681\times 10^{-6}M_{\oplus}$, which is about 50\% of the arcs' mass $M_{\mbox{\scriptsize{arcs}}}$. Because $\Delta EMD_{\mbox{\scriptsize{arcs}}}$ is nearly proportional to $M_{\mbox{\scriptsize{arcs}}}$, as roughly demonstrated by Eq. (\ref{taylor4}), it would become smaller for the lighter arcs. So if we reduce the total mass of the modeled arcs to that of the Smalls, then the two outcomes could match much better as shown in Fig. \ref{fig:validation}(b). More detailed discussion concerning the variation of $M_{\mbox{\scriptsize{arcs}}}$ is made using numerical calculations in Sect. 3.3.

Another important factor in determining $\Delta EMD$ should be the leading-to-trailing number ratio of the Jupiter Trojans, that is, the mass ratio of the arcs. A substantially different $\Delta EMD$ may rise because the asymmetrical variation of the gravitational force experienced by Mars. In the standard model, the two arcs have been assumed to have a ratio of $N[\mbox{L4}]/N[\mbox{L5}]=1.6$, which is very close to that of the Smalls from the observation ($\sim1.7$). This brings little influence of $N[\mbox{L4}]/N[\mbox{L5}]$ on the comparison in Fig. \ref{fig:validation}.

\begin{figure}
  \centering
  \includegraphics[width=9.5cm]{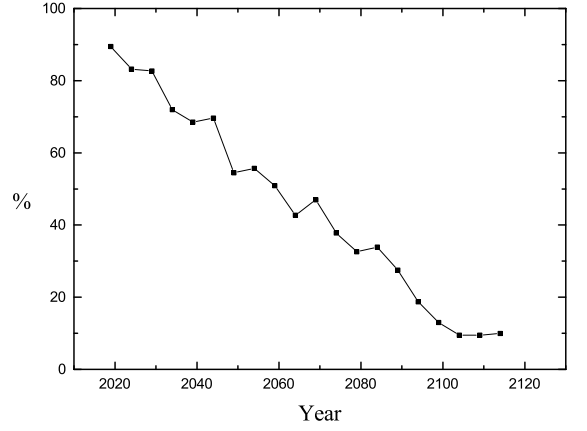}
  \caption{For the case of $M_{\mbox{\scriptsize{arcs}}}=M_{\mbox{\scriptsize{Smalls}}}$ shown in Fig. \ref{fig:validation}(b), the time evolution of the relative difference $R(T)$ (see Eq. (\ref{fig:R(T)})) between the perturbation amplitudes induced by the arcs and the Smalls.}
  \label{fig:R(T)}
\end{figure}

Nevertheless, even the arcs have the mass and number (mass) ratio comparable to the Smalls, in Fig. \ref{fig:validation}(b) there is still a slight distinction in the perturbation amplitudes, defined as the maximum of $|\Delta EMD|$ \footnote{$|\Delta EMD(T)|$ refers to the perturbation amplitude during the time interval 2014--$T$; and $|\Delta EMD|$ refers to the one century case with $T=2114$.}. We use $R(T)$ to denote the relative difference in the amplitude in the interval between years 2014 and $T$, then we have
\begin{equation}
R(T)=\frac{\left|  \mbox{max}|\Delta EMD_{\mbox{\scriptsize{arcs}}}(T)| - \mbox{max}|\Delta EMD_{\mbox{\scriptsize{Smalls}}}(T)| \right|}{\mbox{max}|\Delta EMD_{\mbox{\scriptsize{Smalls}}}(T)|}.
\label{EMD_diff}
\end{equation}
For the case in Fig. \ref{fig:validation}(b), the corresponding time evolution of $R(T)$ is calculated and shown in Fig. \ref{fig:R(T)}. The value of $R(T)$ is very large at the early stage starting from the year 2014, e.g., $R(2019) = 89\%$ over the first five years. Then it decreases gradually and reaches its minimum of $\sim10\%$ near the end of our computation. We find that the capacity of the arcs to model thousands of the Smalls is greatly improved within about 100 years. This result is due to the fact that the arcs are meant to implement the average gravitational effect of the trojan population. Because these objects librate around Jupiter's L4 and L5 points with periods near 150 years (Jewitt et al. 2000), the equivalent effect would not be achieved until they have (nearly) completed their tadpole orbits, in other words, the time interval of computation should be at least on the order of one century. Therefore, we believe that the standard arcs could well represent the global perturbation induced by the small Jupiter Trojans with $H\geqslant11$, for all the observed (i.e., Smalls) and hidden ones.

\subsection{Dependence on the arc parameters}

It is probable that the physical and orbital parameters of Jupiter Trojans will be improved significantly from accurate observations by future space missions. To update planetary ephemerides corresponding to the new data, we should figure out how different parameters related with the trojans would affect our results obtained from the standard arcs. In the following, we explore the dependence of $\Delta EMD_{\mbox{\scriptsize{arcs}}}$ on the arc parameters and evaluate the relative importance of each one.


\subsubsection{Mass}

\begin{figure}
  \centering
  \includegraphics[width=9.5cm]{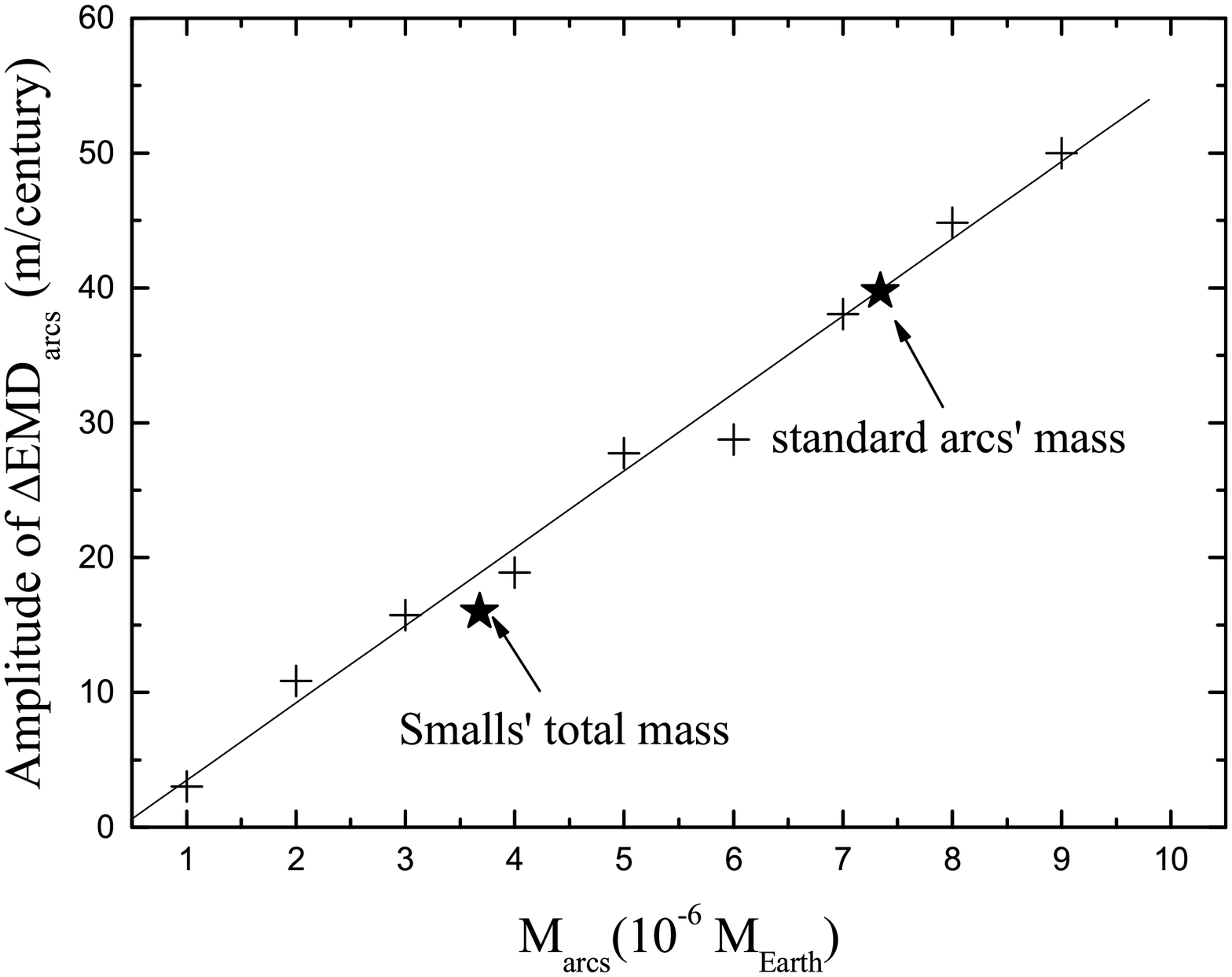}
  \caption{Dependence of the amplitude of $\Delta EMD_{\mbox{\scriptsize{arcs}}}$ on the total mass of the arcs $M_{\mbox{\scriptsize{arcs}}}$. Crosses correspond to the selected values of $M_{\mbox{\scriptsize{arcs}}}$ in the range of $(1-10)\times10^{-6} M_{\oplus}$; while the two stars correspond to the standard mass of $7.343\times10^{-6}M_{\oplus}$ in the arc model, and the Smalls' total mass of $3.681\times10^{-6}M_{\oplus}$, respectively.}
  \label{fig:massPara}
\end{figure}

The largest contributor to the error of $\Delta EMD_{\mbox{\scriptsize{arcs}}}$ should be the uncertainty in the total mass of the Jupiter Trojans, that constrains the value of $M_{\mbox{\scriptsize{arcs}}}$. Therefore, in some extra runs, we chose different $M_{\mbox{\scriptsize{arcs}}}$ in the range of $(1-10)\times10^{-6} M_{\oplus}$, while keeping the other parameters unchanged. 

The results are summarized in Fig. \ref{fig:massPara}. We find that the amplitude of the perturbation on the Earth-Mars distance (crosses) depends linearly on the mass of the arcs, i.e., $|\Delta EMD_{\mbox{\scriptsize{arcs}}}| \propto M_{\mbox{\scriptsize{arcs}}}$. This is consistent with the previous first-order approximation (\ref{taylor4}). Using the linear fitting, we can give a simple description 

\begin{equation}
|\Delta EMD^{(1)}_{\mbox{\scriptsize{arcs}}}|=5.74\cdot \left(\frac{M_{\mbox{\scriptsize{arcs}}}}{10^{-6}M_{\oplus}}\right)-2.26~~\mbox{(m/century)},
\label{EMDmass}
\end{equation}
which is shown as the solid line in Fig. \ref{fig:massPara}. For comparison, we also plot the amplitudes $|\Delta EMD_{\mbox{\scriptsize{arcs}}}|$ corresponding to the total masses of the standard arcs and the observed Smalls. Both of them are indicated by the stars, which are very close to our best fit line. However, if we use Eq. (\ref{EMDmass}) to make an extrapolation, the amplitude $|\Delta EMD_{\mbox{\scriptsize{arcs}}}|$ becomes negative when $M_{\mbox{\scriptsize{arcs}}}\rightarrow0$, this is absolutely incorrect. In practice, it is meaningless to consider the arcs with extremely small mass, which will have a negligible contribution to the improvement of planetary ephemerides. For this reason, here we take $M_{\mbox{\scriptsize{arcs}}}$ starting from $10^{-6} M_{\oplus}$. 

For the arcs, the mass is uniformly distributed in either of one-dimensional azimuthal zones $30^{\circ}-90^{\circ}$ ahead and behind Jupiter's location, so the variation of $M_{\mbox{\scriptsize{arcs}}}$ would not alter their centers of mass, which always remain at Jupiter's L4 and L5 points, respectively. As a result, the arc perturbation does follow a linear relationship to $M_{\mbox{\scriptsize{arcs}}}$ from a physical point of view. But we cannot predict a trend of $\Delta EMD_{\mbox{\scriptsize{Smalls}}}$ increase (decrease) with adding (removing) Smalls, in other words, the change of their total mass. Because in that case, the number (mass) ratio between the L4 and L5 Smalls could vary, and the different spatial (mass) distribution would shift the center of mass of the Smalls around any Lagrangian point. The advantage of our arc approach for the Smalls is that these two factors are no longer coupled with the parameter mass and can be investigated separately.

\subsubsection{Leading-to-trailing number ratio}

\begin{figure}
  \centering
  \includegraphics[width=9.5cm]{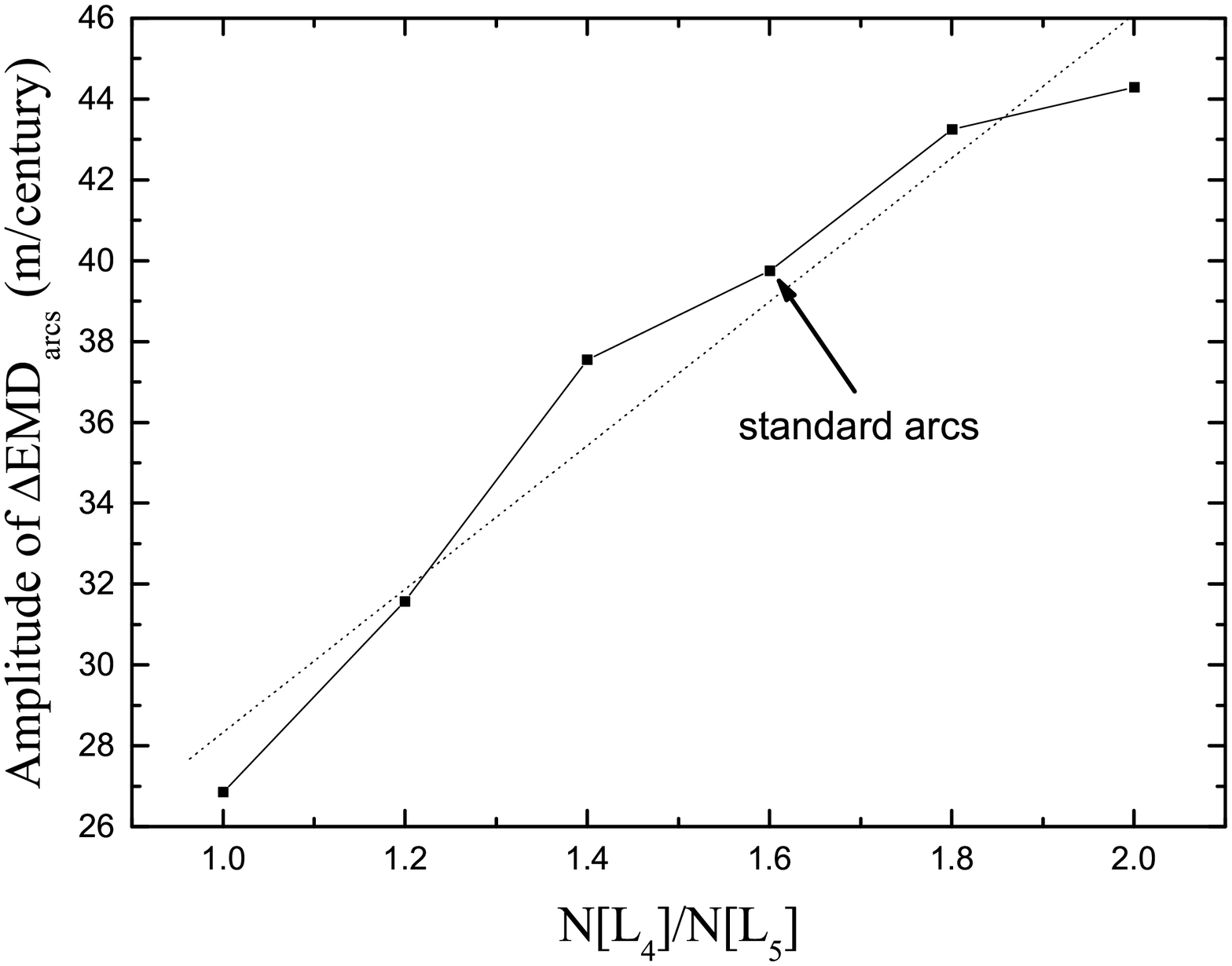}
  \caption{Dependence of the amplitude of $\Delta EMD_{\mbox{\scriptsize{arcs}}}$ on the leading-to-trailing number ratio $N(\mbox{L}4)/N(\mbox{L}5)$ of the Jupiter Trojans with $H\geqslant11$, which is assumed to be equivalent to the mass ratio of the two arcs. For our standard arcs, this ratio is adopted to be 1.6.}
  \label{fig:ratioPara}
\end{figure}

The number ratio of Jupiter Trojans around the leading L4 and trailing L5 points still remains uncertain. Sisto et al. (2014) numerically simulated the long-term evolution of fictitious trojans, and they found that the dynamical stabilities of these two swarms are almost identical. However, for our nominal samples from observations, the numbers of L4 and L5 swarms are 4011 and 2053, indicating a leading-to-trailing number ratio of  $N(\mbox{L}4)/N(\mbox{L}5)=1.95$. This asymmetry could partly come from observational biases, such as the incompleteness at $H>12.3$. Some theoretical works proposed that the unbiased number ratio of trojans could be $N(\mbox{L}4)/N(\mbox{L}5)=1.3$ (Nesvorn\'y et al. 2013), 1.34 (Grav et al. 2012), 1.4 (Grav et al. 2011), and 1.6 (Szab\'o et al. 2007). These values of $N(\mbox{L}4)/N(\mbox{L}5)$ are derived for the entire trojan population, while here we focus on the small trojans with $H\geqslant11$ which are modeled by the arcs. In any case, the leading-to-trailing number ratio always falls in the range of one to two. Thus it seems worthwhile to determine errors in the amplitude of $\Delta EMD_{\mbox{\scriptsize{arcs}}}$ for all the plausible values of N(\mbox{L}4)/N(\mbox{L}5) in such a ratio range. The results are displayed in Fig. \ref{fig:ratioPara}.

Analogous to the investigation of different masses of the arcs reported above, we can simply present the influence of the leading-to-trailing number ratio $N(\mbox{L}4)/N(\mbox{L}5)$ on the perturbation $\Delta EMD_{\mbox{\scriptsize{arcs}}}$. By applying the linear fitting, we have 
\begin{eqnarray}
|\Delta EMD^{(2)}_{\mbox{\scriptsize{arcs}}}|&=&|\Delta EMD^{(1)}_{\mbox{\scriptsize{arcs}}}|+17.7[N(\mbox{L}4)/N(\mbox{L}5)-1.6]\nonumber\\ 
&~&\mbox{(m/century)},
\label{EMDratio}
\end{eqnarray}
where $|\Delta EMD^{(1)}_{\mbox{\scriptsize{arcs}}}|$ is from Eq. (\ref{EMDmass}), and the constant 1.6 is the number ratio used for the standard arcs. The profile of the fitted perturbation amplitude $|\Delta EMD^{(2)}|$ is shown as the dashed line in Fig. \ref{fig:ratioPara}, and it reveals that higher asymmetry in the numbers of leading and trailing Jupiter Trojans leads to stronger perturbation on the Earth-Mars distance.

Based on our results, the currently best-fitted and acceptable range of $N(\mbox{L}4)/N(\mbox{L}5)$ produces the variation of the amplitude of  $\Delta EMD_{\mbox{\scriptsize{arcs}}}$ no larger than $\sim30$\%.  This would not yield the order of magnitude change in the perturbation on the Earth-Mars distance induced by the arcs. Hence we infer that the leading-to-trailing number ratio of Jupiter Trojans has restricted impact on Mars' range, comparing to the influence of the mass.

\subsubsection{Spatial distribution}

\begin{figure}
  \centering
  \includegraphics[width=9.5cm]{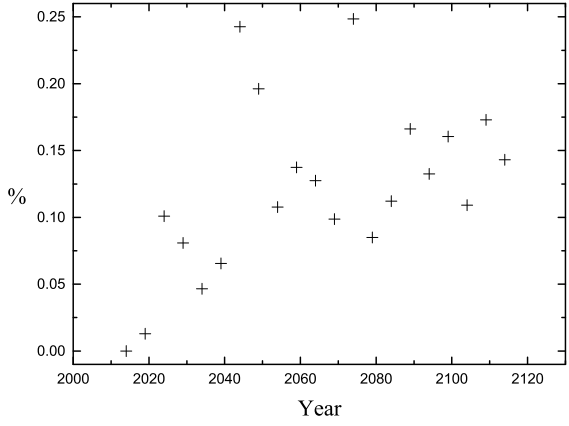}
  \caption{Compared to the standard arcs, given the azimuthal distribution of the arcs similar to that of real Jupiter Trojans, the time evolution of the consequent relative difference $F(T)$ (see Eq. (\ref{EMD_diff})) of the perturbations on the Earth-Mars distance.}
  \label{fig:azimuthPara}
\end{figure}

The azimuthal ($\phi$) distribution of the standard arcs was set to be symmetric respect to the L4/L5 point. While according to the simulated evolution of observed Jupiter Trojans, we find that they could occupy the $\phi$ space from $\phi_1=30^{\circ}$ $(-110^{\circ}$) to $\phi_2=110^{\circ}$ $(-30^{\circ}$) for the L4 (L5) (see the definition in Eq. (\ref{arc})). Let us denote by $F(T)$ the relative difference between the perturbation amplitude $|\Delta EMD^{(3)}_{\mbox{\scriptsize{arcs}}}|$ in such a case and $|\Delta EMD_{\mbox{\scriptsize{arcs}}}|$ for the standard arcs, up to the epoch of year $T$, we have
\begin{equation}
F(T)=\frac{\mbox{max}|\Delta EMD^{(3)}_{\mbox{\scriptsize{arcs}}}(T)| - \mbox{max}|\Delta EMD_{\mbox{\scriptsize{arcs}}}(T)|}{\mbox{max}|\Delta EMD_{\mbox{\scriptsize{arcs}}}(T)|}.
\label{EMD_diff}
\end{equation}

As one can see in Fig. \ref{fig:azimuthPara}, the perturbations caused by the arcs are stronger for the extended distribution of $\phi$, characterized by the index $F(T)>0$. Nevertheless, throughout the calculation over one century, the increment of the amplitude of $\Delta EMD$ is always below 25\%. Since the $\phi$ space adopted here has entirely covered the stable region in azimuth for Jupiter Trojans, not theoretically but in numerical simulations, the difference $F(T)$ should not exceed such an upper limit of 25\%. So we regard the azimuthal distribution of the trojans as another restricted factor in the one-dimensional arc model. 

To improve the total perturbation from the main belt asteroids, recent work by Pitjeva \& Pitjev (2014) takes into account the two-dimensional annulus instead of the one-dimensional ring. They find that the influence of the annulus on the Mars orbit distinguishes it from that of the ring with the same mass, at a level of about 10\%. As for our consideration of the Jupiter Trojans, besides the $\phi$ dimension, we should also evaluate the effect of the mass distribution in the radial ($r$) and vertical ($z$) dimensions. The global perturbation of the small trojans with $H\geqslant11$ is then updated by implementing two three-dimensional uniform arc-columns centered on the L4 and L5 points, with sizes  $r\in[3.8, 6.6\mbox{AU}]$ and $z\in[-3.7, 3.7\mbox{AU}]$ corresponding to the evolution of observed Jupiter Trojans. For the sake of comparison, the other parameters characterizing the arc-columns ($\phi$ distribution, total mass, mass ratio) are the same as those in the set of parameters for the standard arcs.

Numerical experiments have been conducted to calculate the gravitational effect of the three-dimensional arc-columns on the Earth-Mars distance. The resulting perturbation $\Delta EMD$ yields a relative difference of $10\%-25\%$ from the case of one-dimensional arcs during the 2014--2114 time interval. Given that the arc-columns are uniformly distributed in the $r$ and $z$ spaces, while a majority of observed Jupiter Trojans have small eccentricities and low inclinations, suggesting that the bulk of their mass is mainly concentrated around $r=5.2$ AU and $z=0$. So it turns out that we overestimate the difference in the $\Delta EMD$ between the three- and one-dimensional models. Thus, the usage of three-dimensional arc-columns may enhance the ability of modeling perturbations from the trojans, but this improvement seems not significant. Moreover, to fulfill numerical integration of equations of planetary motion simultaneously with perturbations from the arc-columns, much computational time is taken by the three-dimensional integration routine that dramatically delays the code, requiring at least 1000 times longer than the one-dimensional case. It appears to be too computationally expensive in creating the planetary ephemerides, especially when many other factors will have to be introduced, for example, the perturbations from the main belt, Earth tides, Solar quadrupole moment, post relativistic PPN effect, and so on. Based on these facts, we believe that the one-dimensional arc model is more economical, and it should satisfactorily and sufficiently estimate the global perturbation induced by the Jupiter Trojans.   

A further analysis can be made for the spatial distribution of the trojan mass considered here relative to the standard arcs. In the case of the one-dimensional model, the variation of mass distribution in $\phi$ space has moved both centers of mass away from Jupiter's L4 and L5 points, respectively, yielding the change of the perturbation amplitude $|\Delta EMD|$ to a notable degree. But when we expand the one-dimensional standard arcs to three-dimensional arc-columns, the mass distribution remains symmetric with respect to $r=5.2$ AU and $z=0$. Consequently, as long as keeping the density profile in $\phi$ space fixed, the two centers of mass would not be displaced at all, and the value of $|\Delta EMD|$ should not increase or decrease by such an extent over $10\%$. We believe the reason is that the reference plane (i.e., $z=0$) for our numerical calculations is the ecliptic, which has a small tile ($<2^\circ$) with respect to the orbital plane of Earth or Mars. When the mass has a distribution in the $z$ space, even uniform and symmetric, the perturbation on the Earth-Mars distance from the potential $U(z>0)$ is not exactly equal to that from $U(z<0)$, leading to some fluctuations but not a consistent trend in the difference of $|\Delta EMD|$.

\subsection{Overall result} 

Combining the result from the standard arc model in Sect. 3.1 with the analysis of the dependency of model parameters on the observations in Sect. 3.3, we finally obtain a simple fitting formula for the global perturbation induced by the Jupiter Trojans on the Earth-Mars distance

\begin{eqnarray}
|\Delta EMD|&=&30 +\alpha\cdot\beta\left[5.74 \left(\frac{M_{\mbox{\scriptsize{arcs}}}}{10^{-6}M_{\oplus}}\right)+17.7\frac{N(\mbox{L}4)} {N(\mbox{L}5)} - 30.69 \right]\nonumber\\
&~&\mbox{(m/century)},
\label{EMDfinal}
\end{eqnarray}
where the first term on the right side is contributed by the Bigs, and the second term is from the remaining small trojans with absolute magnitudes $H\geqslant11$, including those have not yet been discovered. 

In Eq. (\ref{EMDfinal}), the total mass of the arcs $M_{\mbox{\scriptsize{arcs}}}$, equivalent to that of the small trojan population with $H\geqslant11$, is the most important and uncertain parameter, and it will have to be better determined by future observations. The leading-to-trailing number ratio $N(\mbox{L}4)/N(\mbox{L}5)$ of the Jupiter Trojans is well constrained in the range of one to two. As for the constant $\alpha$ concerning the spatial distribution of the trojans in the $\phi$ space, and the constant $\beta$ for the $r$ and $z$ spaces, both of them could possibly vary within the range of 1--1.25. Here, for the mass variation in any space, we merely present the corresponding upper limit of the difference in $\Delta EMD$ from the standard arcs. Further attention could be paid to different density profiles and spatial scales.


\section{Application to satellite populations}

Having constructed the new arc model of the Jupiter Trojans for the development of planetary ephemerides, we now apply it to the motion of Martian and Jovian satellites. As an additional contribution in the gravitational field of the planet-satellite system, the perturbations are from the 226 Bigs and the two standard one-dimensional arcs. By utilizing a similar procedure for the evaluation of $\Delta EMD$, we can easily estimate the change in the Earth-satellite distance, which is denoted by $\Delta ESD$ in the following. Rather than confined to the ephemerides of planets, investigation of this aspect would make our arc model more useful.

\subsection{Perturbation on Martian satellites}

The position vector of a Martian satellite with respect to Earth can be written as
\begin{equation}
\vec{r}_{\mbox{\scriptsize{ES}}}=\vec{r}_{\mbox{\scriptsize{EM}}}+\vec{r}_{\mbox{\scriptsize{MS}}}, 
\label{vector}
\end{equation}
where $\vec{r}_{\mbox{\scriptsize{EM}}}$ denotes the position vector of Mars relative to Earth, and $\vec{r}_{\mbox{\scriptsize{MS}}}$ is the position vector of the satellite to Mars. 

The perturbation of the Jupiter Trojans has, in essence, modified the angular velocity of Mars ($\omega_{\mbox{\scriptsize{Mars}}}$). Consequently, this changes the phase angle between Mars and Earth from the perspective of the Sun, leading to the variation of the vector $\vec{r}_{\mbox{\scriptsize{EM}}}$. Because of the rather large mass of the Mars, the gravitational force from the Jupiter Trojans is too weak to affect $\omega_{\mbox{\scriptsize{Mars}}}$ significantly on a short timescale, and we found that the change of the modulus $|\vec{r}_{\mbox{\scriptsize{EM}}}|$ (i.e., $\Delta EMD$) is only about 70 m per century. But regarding the Martian satellites, as they are much smaller, the same trojan force could apply considerable accelerations to their angular velocities. Thus we suppose that the impact of the Jupiter Trojans on the motions of these moons around Mars should be much larger, and the difference in the vector $\vec{r}_{\mbox{\scriptsize{MS}}}$ has an upper bound of $2|\vec{r}_{\mbox{\scriptsize{MS}}}|$. Therefore, as the sum of vectors $\vec{r}_{\mbox{\scriptsize{EM}}}$ and $\vec{r}_{\mbox{\scriptsize{MS}}}$, $\vec{r}_{\mbox{\scriptsize{ES}}}$ could also experience a significant change, which is equivalent to a large value of $\Delta ESD$.


\begin{figure}
  \centering
  \begin{minipage}[c]{0.5\textwidth}
  \centering
  \hspace{0cm}
  \includegraphics[width=9.5cm]{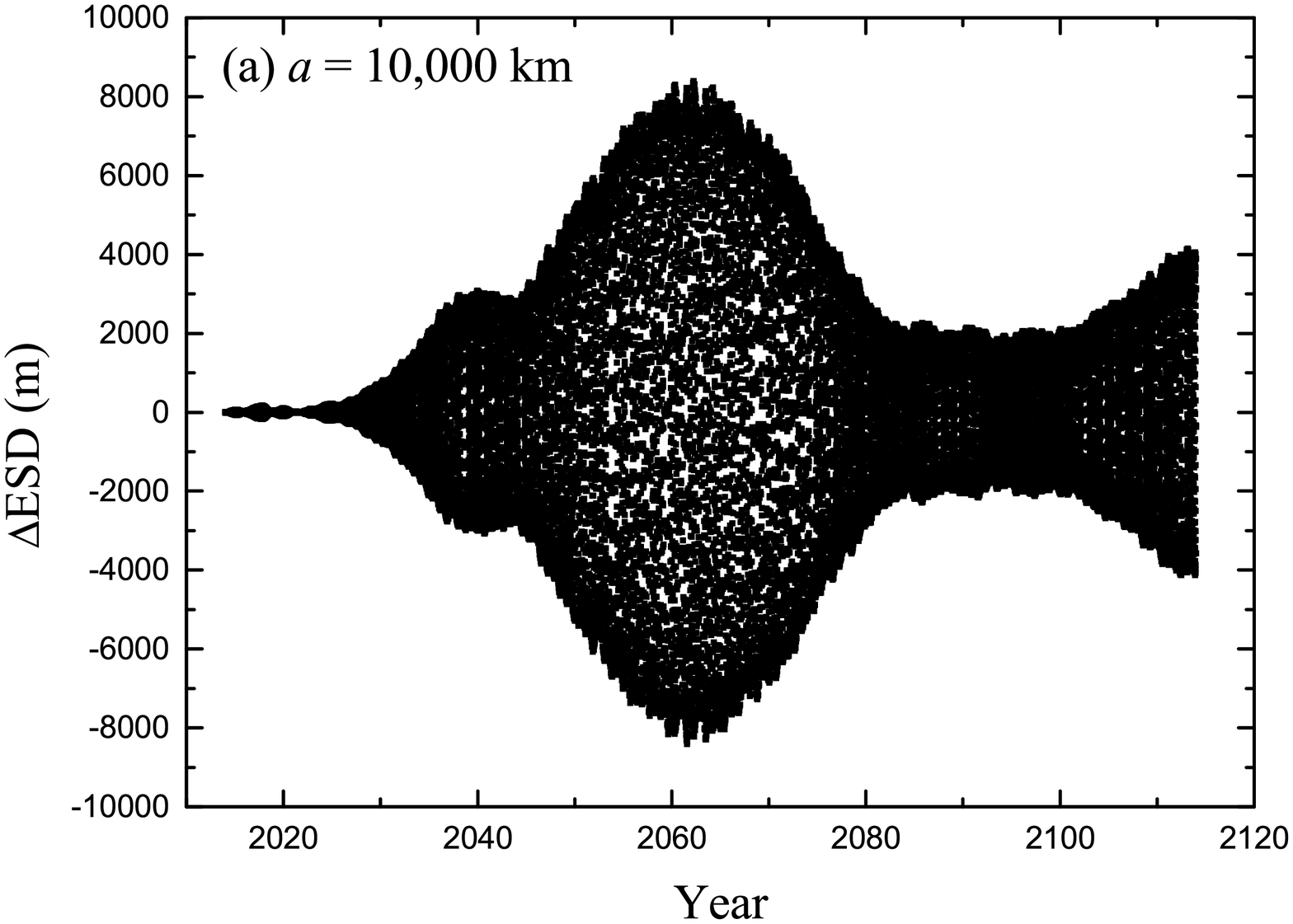}
  \end{minipage}
  \begin{minipage}[c]{0.5\textwidth}
  \centering
  \hspace{0cm}
  \includegraphics[width=9.5cm]{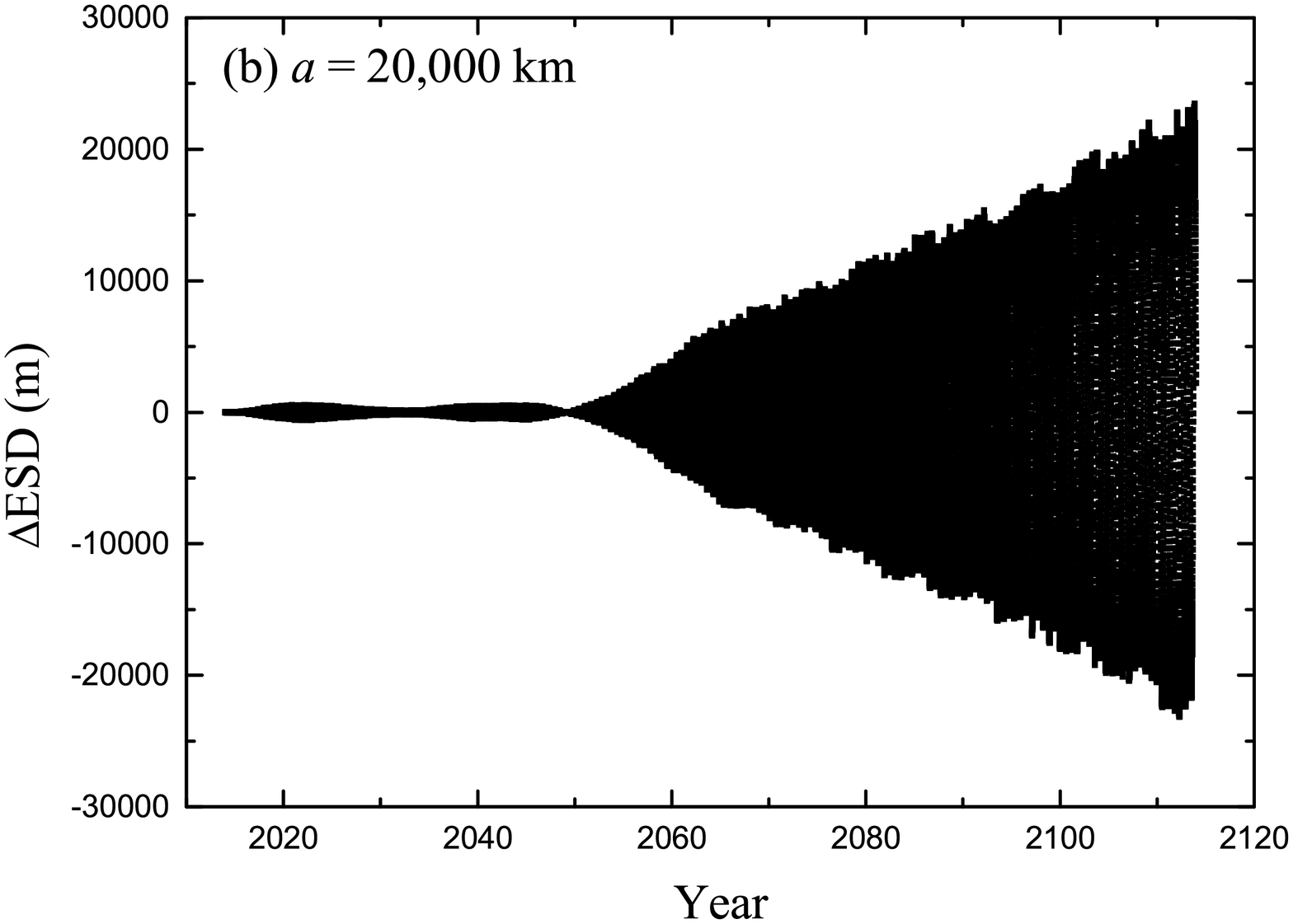}
  \end{minipage}
  \caption{Perturbation induced on the distance between Earth and a Martian pesudo-satellite by the Jupiter Trojans: (a) Phobos-like satellite with semimajor axis $a=$ 10,000 km; (b) Deimos-like satellite with $a=$ 20,000 km.}
 \label{fig:MarSat}
\end{figure}  

The two natural satellites of Mars are Phobos and Deimos, which orbit Mars with semimajor axes of 9,377 km and 23,460 km, respectively. Taking into consideration that the current and future Martian probes could also reach such heights, for these natural and artificial satellites revolving around Mars, we have adopted two representative samples with semimajor axes $a=10,000$ km and $20,000$ km on near-circular orbits close to the equatorial plane of Mars. Numerical integrations are then performed to evaluate the perturbing effects of the Jupiter Trojans, in the framework of the standard arc model.

Figure \ref{fig:MarSat} shows the time evolution of $\Delta ESD$ for the two pesudo-satellites. The amplitudes of perturbations with the maximum can reach on the order of 10,000 m per century, which are two orders of magnitude larger than that of $\Delta EMD$. The numerical results agree very well with our analytic prediction at the beginning of this subsection. And the arc model may enable us to calculate more accurate perturbations on Martian satellites. Furthermore, apart from semimajor axis, eccentricity and inclination, the position of a Martian satellite relative to Earth also depends on the other three orbital elements: mean longitude, longitude of pericenter, and longitude of ascending node. These three angles determine the initial phase of the satellite, which can drive substantial oscillations of the direction of the vector $\vec{r}_{\mbox{\scriptsize{MS}}}$ and induce quite different $\vec{r}_{\mbox{\scriptsize{ES}}}$. As can be seen, this argument explains why the evolutionary behaviors of $\Delta ESD$ in Fig. \ref{fig:MarSat}(a) and (b) are so distinct on the 2014--2114 time interval. Eventually, the amplitude of $\Delta ESD$ will increase gradually as time continues to pass.

Regardless of the influence of the initial phases of the Martian satellites, we note that the perturbation $\Delta ESD$ induced by the Jupiter Trojans could always be suppressed for a while at the early stage of our integrations. For instance, as shown in Fig. \ref{fig:MarSat}(a), the amplitude of $\Delta ESD$ can sustain a small value of $\sim$100--200 m for the inner sample, similar to Phobos. This feature makes it plausible that the perturbation of the Jupiter Trojans is not a dominant factor for the Mars Orbiters on a short timescale, such as within ten years. 

\subsection{Perturbation on Jovian satellites}

\begin{figure}
  \centering
  \begin{minipage}[c]{0.5\textwidth}
  \centering
  \hspace{0cm}
  \includegraphics[width=9.5cm]{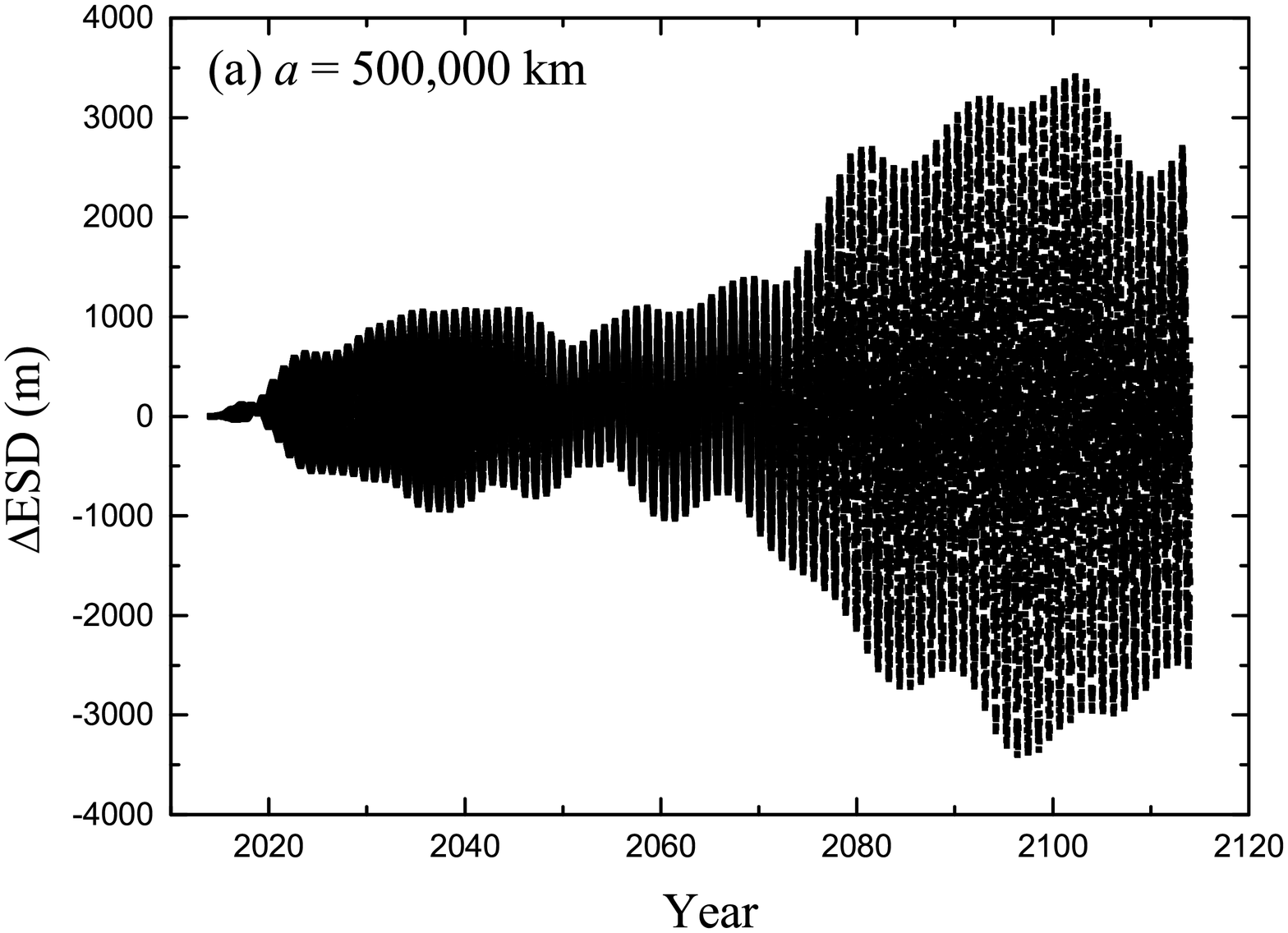}
  \end{minipage}
  \begin{minipage}[c]{0.5\textwidth}
  \centering
  \hspace{0cm}
  \includegraphics[width=9.5cm]{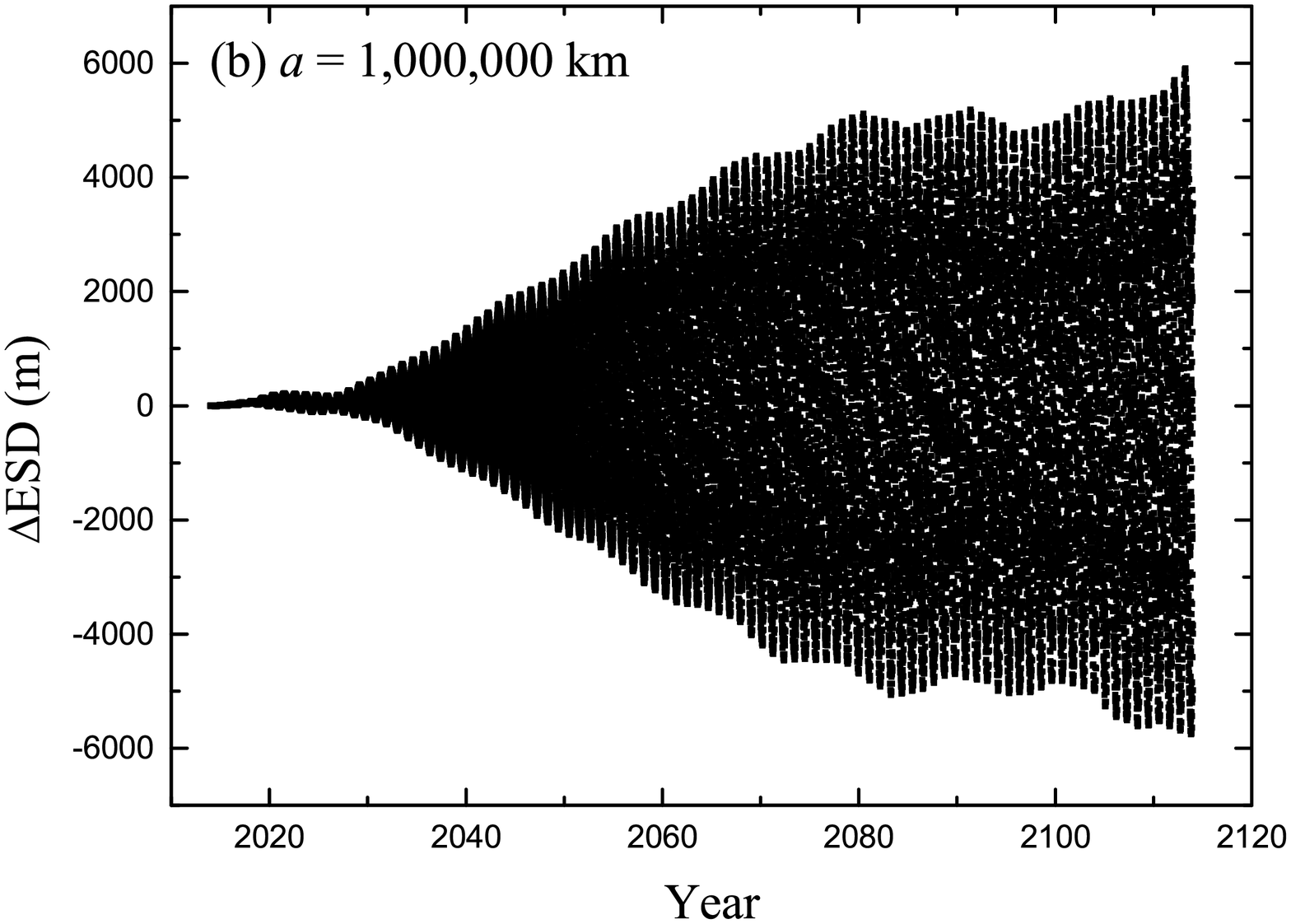}
  \end{minipage}
  \begin{minipage}[c]{0.5\textwidth}
  \centering
  \hspace{0cm}
  \includegraphics[width=9.5cm]{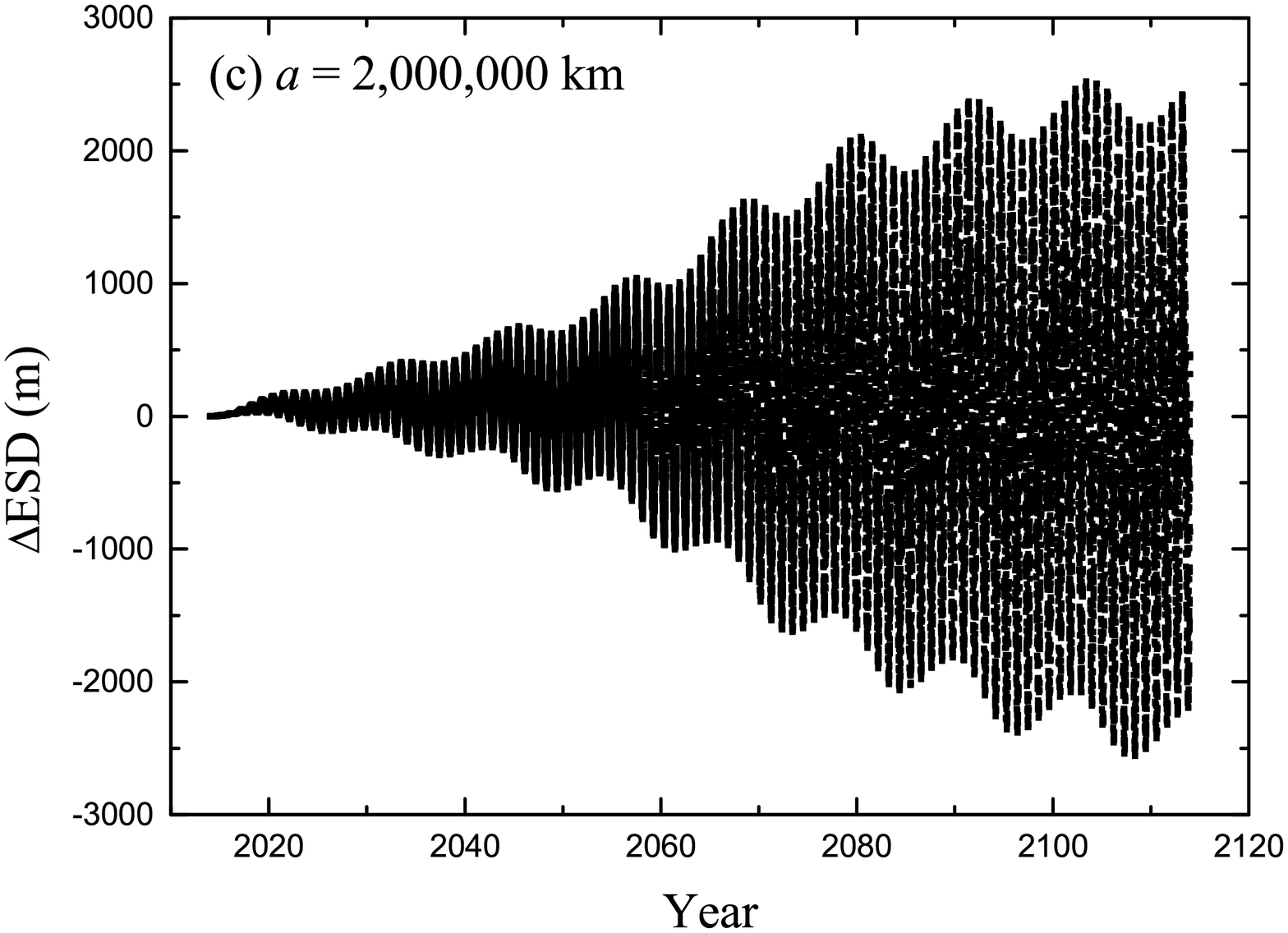}
  \end{minipage}
  \caption{As in Fig. \ref{fig:MarSat}, but for three representative Jovian pesudo-satellites with: (a) $a=$ 500,000 km; (b) $a=$ 1,000,000 km; (c) $a=$ 2,000,000 km.}
 \label{fig:JovSat}
\end{figure}  

Next we turn to investigate the satellites of Jupiter listed on the JPL's website \footnote{https://ssd.jpl.nasa.gov/?sat\_elem}. For the four largest satellites, they are moving on the orbits with semimajor axes of 421,800 km (Io), 671,100 km (Europa), 1,070,400 km (Ganymede), 1,882,700 km (Callisto); and, all of them have very small eccentricities ($<0.01$) and low inclinations ($\lesssim2^\circ$). Analogously, we did the same for representative orbits with semimajor axes of $a=$ 500,000 km, 1,000,000 km, 2,000,000 km, to roughly evaluate the orbital deviations of Jovian satellites caused by their host planet's trojans. The measurements of $\Delta ESD$ are shown in Fig. \ref{fig:JovSat}. 

The typical perturbation $\Delta ESD$ experienced by the three Jovian pesudo-satellites is on the order of thousands of meters during the time interval 2014--2114. Such $\Delta ESD$ is a bit smaller but still comparable to that of the Martian satellites. Although the four considered satellites have rather small angular velocities due to their large $a$, they are actually far away from any swarm of Jupiter Trojans by about 5.2 AU. The consequent weaker accelerations induced by the trojan population bring down the net changes in their angular velocities, as does $\Delta ESD$.


\section{Conclusions}

In this paper, we have built a new arc model to represent the global perturbation induced by the Jupiter Trojans for the purpose of refining the dynamical model of the solar system and developing the numerical planetary ephemerides. The population of Jupiter Trojans have been divided into two subgroups: (1) the Bigs, 226 objects with absolute magnitudes $H<11$. They are not only observational complete but also have well determined masses; (2) the remaining small objects with $H\ge11$. They are far from observational completeness and most have no available diameter data. By comparing the observed samples with the prediction from collisional equilibrium theory (Kenyon \& Bromley 2004, 2012), for these fainter trojans, we firstly determined their absolute magnitude distribution all the way to $H=\infty$; and then their total mass is estimated to be about $7.343\times10^{-6}M_{\oplus}$. We note that this value includes masses of those small trojans which have not been discovered yet. Adding the masses of the Bigs, we can therefore derive that the total mass of the Jupiter Trojans is most likely $\sim1.861\times 10^{-5}M_{\oplus}$.

Usage of the obtained trojan masses has allowed to evaluate the perturbations of the Jupiter Trojans: the 226 Bigs are treated as individuals, and the remaining small trojans are modeled by two discrete massive arcs centered at Jupiter's L4 and L5 points. We then performed integrations of the equations of planetary motion in the perturbed and unperturbed solar systems, respectively. The results indicate that the total effect of the Jupiter Trojans on the change in the Earth-Mars distance reaches on a value of $\sim$70 m during the 2014--2114 time interval. The uncertainties of several parameters characterizing the two arcs have also been taken into account, and the influences are summarized neatly in Eq. (\ref{EMDfinal}).


As applications of the arc model in the solar system, we further investigated the perturbations on Martian and Jovian satellites induced by the Jupiter Trojans. Since these satellites have rather small masses, their motions can be affected to a significant extent. According to our estimations, the change of the distance between Earth and a Martian or Jovian satellite could be as large as tens of thousands meters over a century. 

\begin{acknowledgements}
This work was supported by the National Natural Science Foundation of China (Nos. 11473015, 11333002 and 11178006). We are grateful to Prof. Yanning Fu for helpful discussions. The authors would also like to express their thanks to the referee for her valuable comments.

\end{acknowledgements}


\begin{thebibliography}{}

\bibitem[2014]{romi14} Di Sisto, R. P., Ramos, X. S., \& Beaug\'e, C. 2014, Icarus, 243, 287

\bibitem[2003]{fern03} Fern\'andez, Y. R., Sheppard, S. S., \& Jewitt, D. C. 2003, AJ, 126, 1563

\bibitem[2009]{fern09} Fern\'andez, Y. R., Jewitt, D., \& Ziffer, J. E. 2009, AJ, 138, 240

\bibitem[2008]{fien08} Fienga, A., Manche, H., Laskar, J., \& Gastineau, M. 2008, A\&A, 477, 315

\bibitem[2010]{fien10} Fienga, A., Manche, H., Kuchynka, P., et al. 2010 [arXiv:1011.4419]

\bibitem[2011]{fien11} Fienga, A., Laskar, J., Kuchynka, P., et al. 2011, Celestial Mechanics and Dynamical Astronomy, 111, 363

\bibitem[2008]{folk08} Folkner, W. M., Williams, J. G., \& Boggs, D. H. 2008, JPL IOM 343R-08-003

\bibitem[2011]{grav11} Grav, T., Mainzer, A. K., Bauer, J., et al. 2011, ApJ, 742, 40

\bibitem[2012]{grav12} Grav, T., Mainzer, A. K., Bauer, J. M., et al. 2012, ApJ, 759, 49

\bibitem[2000]{jewi00} Jewitt, D. C., Trujillo C. A., \& Luu J. X. 2000, AJ, 120, 1140

\bibitem[2004]{jewi04} Jewitt, D. C., Sheppard, S., \& Porco, C. 2004, in Jupiter: The Planet, Satellites and Magnetosphere, ed. F. Bagenal, T. Dowling, \& W. McKinnon (Cambridge: Cambridge Univ. Press), 263

\bibitem[2004]{keny04} Kenyon, S. J., \& Bromley, B. C. 2004. AJ, 128, 1916

\bibitem[2012]{keny12} Kenyon, S. J., \& Bromley, B. C. 2012. ApJ, 143, 63

\bibitem[2006]{kono06} Konopliv, A. S., Yoder, C. F., Standish, E. M., et al. 2006, Icarus, 182, 23

\bibitem[2002]{kras02} Krasinsky, G. A., Pitjeva, E. V., Vasilyev, M. V., \& Yagudina, E. I. 2002, Icarus, 158, 98

\bibitem[2011]{kono11}  Krasinsky, G. A., Asmar, S. W.,  Folkner, W. M., et al. 2011, Icarus,  211, 401

\bibitem[2010]{kuch10} Kuchynka, P., Laskar, J.,  Fienga, A., \& Manche, H. 2010, A\&A, 514, A96

\bibitem[2005]{morb05} Morbidelli, A., Levison, H. F., Tsiganis, K., \& Gomes, R. 2005, Nature, 435, 462

\bibitem[2009]{morb09} Morbidelli, A., Levison, H. F., Bottke, W. F., et al. 2009, Icarus, 202, 310

\bibitem[2008]{naka08} Nakamura, T., \& Yoshida, F. 2008, PASJ, 60, 293

\bibitem[2013]{nesv13} Nesvorn\'y, D., Vokrouhlick\'y, D., \& Morbidelli, A. 2013, ApJ, 768, 45

\bibitem[2014]{pitj14} Pitjeva, E. V., \& Pitjev, N. P.  2014, Celestial Mechanics and Dynamical Astronomy, 119, 237

\bibitem[1995]{stan95} Standish, E. M., 1998, JPL, IOM 312.F-98-048 
                                            (http://ssd.jpl.nasa.gov/iau-comm4/de405iom/)
                         
\bibitem[2007]{szab07} Szab\'o, Gy. M.,  Ivezi\'c, \v{Z}., Juri\'c, M., \& Lupton, R. 2007, MNRAS, 377, 1393

\bibitem[2015]{vino15} Vinogradova, T. A. \& Chernetenko, Yu. A. 2015, Solar System Research, 49, 391

\bibitem[2008]{yosh08} Yoshida, F., \& Nakamura, T. 2008, PASJ, 60, 297
                                                                                                                 
\end{thebibliography}
\end{document}